\documentclass[aps,twocolumn,prl]{revtex4-1}
\usepackage{CJK}
\usepackage{lipsum}
\usepackage{epsfig,amsmath,amssymb}
\usepackage{color}
\usepackage{float}
\usepackage{subfigure}
\usepackage[utf8]{inputenc}
\usepackage{amsfonts}
\usepackage{amsmath}
\usepackage{hyperref}
\usepackage{amssymb}
\usepackage[english]{babel}
\usepackage{graphicx}
\usepackage{epsfig}
\usepackage{bm}
\usepackage{longtable}
\usepackage{verbatim}
\usepackage{longtable}
\usepackage{lipsum}
\definecolor{BrickRed}{rgb}{0.9,0.1,0}

\newcommand{\mathsym}[1]{{}}

\newcommand{\hs}{\hspace*{0.3cm}}

\newcommand{\be}{\begin{equation}}
\newcommand{\ee}{\end{equation}}
\newcommand{\bea}{\begin{eqnarray}}
\newcommand{\eea}{\end{eqnarray}}
\newcommand{\ben}{\begin{enumerate}}
\newcommand{\een}{\end{enumerate}}
\newcommand{\bit}{\begin{itemize}}
\newcommand{\eit}{\end{itemize}}
\newcommand{\bde}{\begin{widetext}}
\newcommand{\ede}{\end{widetext}}
\newcommand{\nn}{\nonumber}
\newcommand{\crn}{\nonumber \\}

\newcommand{\al}{\alpha}
\newcommand{\la}{\lambda}

\newcommand{\va}{\varphi}

\newcommand{\fr}{\frac}

\newcommand{\bc}{\begin{center}}
\newcommand{\ec}{\end{center}}
\newcommand{\Ga}{\Gamma}
\newcommand{\de}{\delta}
\newcommand{\De}{\Delta}
\newcommand{\ep}{\epsilon}

\newcommand{\La}{\Lambda}
\newcommand{\si}{\sigma}

\newcommand{\Om}{\Omega}

\input{tcilatex}
\begin{document}

\title{Minimal model for the fermion flavor structure, mass hierarchy, dark matter, leptogenesis, and the electron and muon anomalous magnetic moments}
\author{A. E. C\'arcamo Hern\'andez$^a$}
\email{antonio.carcamo@usm.cl}
\author{D. T. Huong$^{b}$}
\email{dthuong@iop.vast.ac.vn}
\author{H. N. Long$^{c,d}$}
\email{hoangngoclong@tdtu.edu.vn}
\affiliation{$^a$Departamento de F\'{\i}sica, Universidad T\'{e}cnica Federico Santa Mar\'{\i}a,\\
Casilla 110-V, Valpara\'{\i}so, Chile\\
$^b$ Institute of Physics, VAST, 10 Dao Tan, Ba Dinh, Hanoi, Vietnam\\
$^c$ Theoretical Particle Physics and Cosmology Research Group, Advanced
Institute of Materials Science, Ton Duc Thang University, Ho Chi Minh City,
Vietnam\\
$^d$ Faculty of Applied Sciences, Ton Duc Thang University, Ho Chi Minh City, Vietnam}
\date{\today}

\begin{abstract}
We propose a renormalizable theory with minimal particle content and symmetries, that successfully explains the number of Standard Model (SM) fermion families, the SM fermion mass hierarchy, the tiny values for the light active neutrino masses, the lepton and baryon asymmetry of the Universe, the dark matter relic density as well as the muon and electron anomalous magnetic moments. In the proposed model, the top quark and the exotic fermions do acquire tree-level masses whereas the SM charged fermions lighter than the top quark gain one-loop level masses. Besides that, the tiny masses for the light active neutrino are generated from an inverse seesaw mechanism at one-loop level.

\footnotesize
DOI:\href{https://doi.org/10.1103/PhysRevD.102.055002}{10.1103/PhysRevD.102.055002}
\normalsize
\end{abstract}

\maketitle

\begin{figure}[th]
\includegraphics[width=0.9\textwidth]{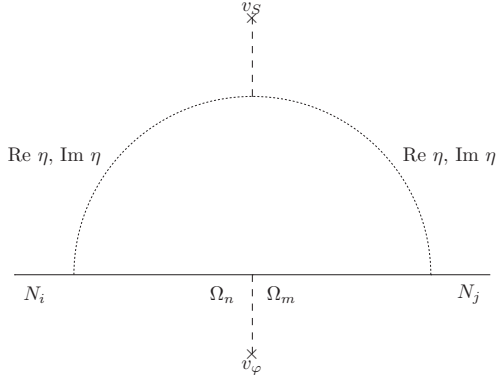}
\vspace{-14cm}
\caption{One-loop Feynman diagram contributing to the Majorana neutrino mass submatrix $\protect\mu$. Here, $n=1,2$, $i,j=1,2,3$.}
\label{Loopdiagramsmu}
\end{figure}

\begin{figure}[th]
\resizebox{9.0cm}{15cm}{\vspace{-2cm}%
\includegraphics{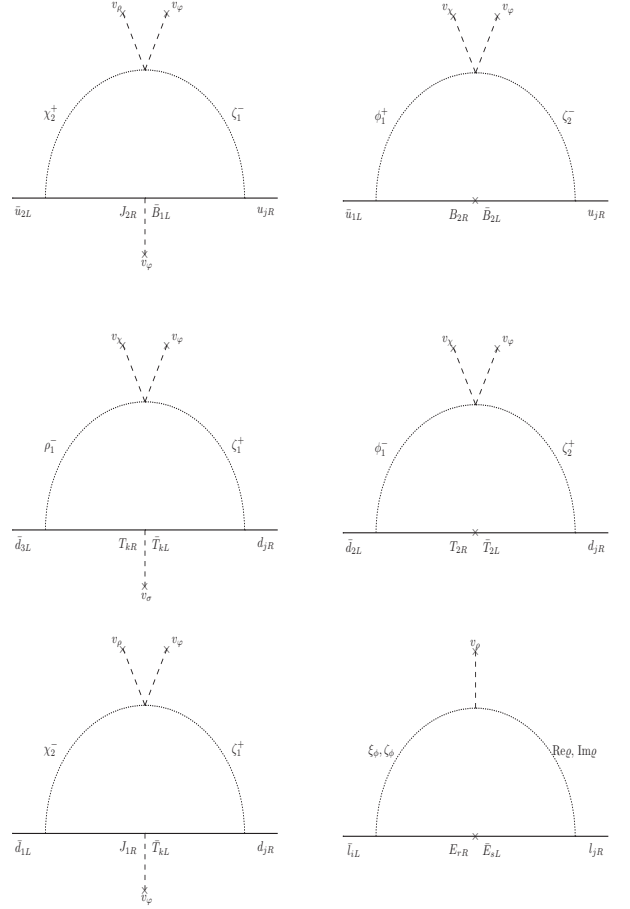}}
\caption{One-loop Feynman diagrams contributing to the entries of the SM
charged fermion mass matrices. Here, $n=1,2$, $k=1,3$ and $i,j,r,s=1,2,3$.}
\label{Loopdiagramsf}
\end{figure}

Despite the considerable experimental success of the Standard Model (SM), it leaves many unanswered questions such the origin of the observed SM fermion mass hierarchy, the number of SM fermion families, the tiny values of the light active neutrino masses, the nature of the dark matter (DM), the electric charge quantization, the lepton and baryon asymmetries of the Universe and the anomalous magnetic moments of the muon and electron. To address these issues,
we propose a minimal renormalizable theory with the extended $%
SU(3)_C\times SU(3)_L\times U(1)_X$ gauge symmetry, which is
supplemented by the $U(1)_{L_g}$ lepton number symmetry and the $Z_4$
discrete group. Its scalar and fermionic spectrum with their assignments
under the symmetries of the model are shown in Tables \ref{tab:scalars} and %
\ref{tab:fermions}, respectively. The above gauge symmetry is crucial for explaining the number of SM fermion
families since to fulfill the anomaly cancellation
conditions, the number of left-handed $SU(3)_L$ fermion triplets has to be
equal to the number of $SU(3)_L$ fermion antitriplets, which only happens
when the number of fermion generations is a multiple of three. Additionally, the charge quantization in theories with 
$SU(3)_C\times SU(3)_L\times U(1)_X$ gauge symmetry is explained thanks to the anomaly cancellation as shown for the 
first time in \cite{VanDong:2005ux}. The $%
U(1)_{L_g}$ lepton number symmetry is spontaneously broken down to the
residual $Z_2^{(L_g)}$ lepton number symmetry, under which only leptons
are charged, thus forbidding proton decay \cite{CarcamoHernandez:2017cwi}. In this model, the lepton number is defined as:
\be
L=\fr{4}{\sqrt{3}}T_8+L_g,
\ee%
where $L_g$ is a conserved charge associated with the $U(1)_{L_g}$
global lepton number symmetry. Furthermore, the $Z_4$ discrete group is spontaneously broken down to the preserved $Z_2$
symmetry, which allows stable DM candidates that trigger a one-loop level radiative inverse seesaw mechanism for the generation of the light
active neutrino masses. Such DM candidates also mediate the one-loop level radiative seesaw mechanism that produces the SM charged lepton
masses.
\begin{table}[th]
\begin{tabular}{|c|c|c|c|c|c|c|c|c|c|c|}
\hline
& $\chi $ & $\rho $ & $\phi $ & $\si  $ & $\va$ & $\varrho $ & $\eta $
& $S$ & $\zeta_1^\pm $ & $\zeta _2^\pm $ \\ \hline
$SU(3)_C$ & $\mathbf{1}$ & $\mathbf{1}$ & $\mathbf{1}$ & $\mathbf{1}$ & $%
\mathbf{1}$ & $\mathbf{1}$ & $\mathbf{1}$ & $1$ & $\mathbf{1}$ & $\mathbf{1}$
\\ \hline
$SU(3)_L$ & $\mathbf{3}$ & $\mathbf{3}$ & $\mathbf{3}$ & $\mathbf{1}$ & $%
\mathbf{1}$ & $\mathbf{1}$ & $\mathbf{1}$ & $1$ & $\mathbf{1}$ & $\mathbf{1}$
\\ \hline
$U(1)_X$ & $-\fr 1 3$ & $\fr 2 3$ & $\fr 2 3$ & $0$ & $0$ & $0$ & $0$
& $0$ & $\pm 1$ & $\pm 1$ \\ \hline
$U(1)_{L_g}$ & $\fr 4 3$ & $-\fr 2 3$ & $-\fr 2 3$ & $0$ & $2$ & $0$ &
$2$ & $4$ & $0$ & $0$ \\ \hline
$Z_4$ & $1$ & $1$ & $i$ & $\mathbf{-}1$ & $-1$ & $-i$ & $-i$ & $-1$ & $-1$ &
$-i$ \\ \hline
\end{tabular}%
\caption{Scalar assignments under $SU(3)_C\times SU(3)_L\times\times U(1)_X\times U(1)_{L_g}\times Z_4$.}
\label{tab:scalars}
\end{table}
\begin{widetext}
\begin{table}[th]
\renewcommand{\arraystretch}{1.2}
\begin{tabular}{|c|c|c|c|c|c|c|c|c|c|c|c|c|c|c|c|c|c|c|c|c|c|c|}
\hline
& $Q_{1L}$ & $Q_{2L}$ & $Q_{3L}$ & $u_{iR}$ & $d_{iR}$ & $J_{1R}$ & $J_{2R}$
& $J_{3R}$ & $T_{kL}$ & $T_{kR}$ & $T_{2L}$ & $T_{2R}$ & $B_{1L}$ & $B_{1R}$
& $B_{2L}$ & $B_{2R}$ & $L_{iL}$ & $l_{iR}$ & $E_{iL}$ & $E_{iR}$ & $N_{iR}$
& $\Om_{nR}$ \\ \hline
$SU(3)_C$ & $\mathbf{3}$ & $\mathbf{3}$ & $\mathbf{3}$ & $\mathbf{3}$ & $%
\mathbf{3}$ & $\mathbf{3}$ & $\mathbf{3}$ & $\mathbf{3}$ & $\mathbf{3}$ & $%
\mathbf{3}$ & $\mathbf{3}$ & $\mathbf{3} $ & $\mathbf{3}$ & $\mathbf{3}$ & $%
\mathbf{3}$ & $\mathbf{3}$ & $\mathbf{1}$ & $\mathbf{1}$ & $\mathbf{1}$ & $%
\mathbf{1}$ & $\mathbf{1}$ & $\mathbf{1}$ \\ \hline
$SU(3)_L$ & $\mathbf{3}$ & $\mathbf{\overline{3}} $ & $\mathbf{\overline{3}%
} $ & $\mathbf{1}$ & $\mathbf{1}$ & $\mathbf{1}$ & $\mathbf{1}$ & $\mathbf{1}
$ & $\mathbf{1}$ & $\mathbf{1}$ & $\mathbf{1}$ & $\mathbf{1}$ & $\mathbf{1}$
& $\mathbf{1}$ & $\mathbf{1}$ & $\mathbf{1} $ & $\mathbf{3}$ & $\mathbf{1}$
& $\mathbf{1}$ & $\mathbf{1}$ & $\mathbf{1}$ & $\mathbf{1}$ \\ \hline
$U(1)_X$ & $\fr 1 3$ & $0$ & $0$ & $\fr 2 3$ & $-\fr 1 3$ & $%
\fr 2 3$ & $-\fr 1 3 $ & $-\fr 1 3$ & $\fr 2 3$ & $\fr 2 3
$ & $\fr 2 3$ & $\fr 2 3$ & $-\fr 1 3$ & $-\fr 1 3$ & $-%
\fr 1 3$ & $-\fr 1 3$ & $-\fr 1 3$ & $-1$ & $-1$ & $-1$ & $0$ & $%
0$ \\ \hline
$U(1)_{L_g}$ & $-\fr 2 3$ & $\fr 2 3$ & $\fr 2 3$ & $0$ & $0$
& $-2$ & $2$ & $2$ & $0$ & $0$ & $0$ & $0$ & $0$ & $0$ & $0$ & $0$ & $\fr 1 3$
 & $1$ & $1$ & $1$ & $-1$ & $-1$ \\ \hline
$Z_4$ & $-1$ & $-1$ & $1$ & $1$ & $1$ & $-1$ & $-1$ & $1$ & $-1$ & $1$ & $%
-i$ & $-i$ & $1$ & $-1$ & $i$ & $i$ & $1$ & $-1$ & $i$ & $-i$ & $1$ & $i$ \\
\hline
\end{tabular}%
\caption{Fermion assignments under $SU(3)_C\times SU(3)_L\times U(1)_X \times U(1)_{L_g}\times Z_4$. Here $n=1,2$, $k=1,3$
and $i=1,2,3$. Notice that there are six $SU(3)_L$ fermionic triplets (three from $Q_{1L}$ and three from $L_{iL}$)
 and six $SU(3)_L$ fermionic antitriplets, i.e., $Q_{nL}$ ($n=1,2$). In this counting the color quantum numbers are taken into account.}
\label{tab:fermions}
\end{table}
\end{widetext}
\vspace{0cm}
Notice, as follows from Table \ref{tab:scalars} that the
electrically neutral gauge singlet scalars $\eta $ and $\varrho $ as well as
the neutral components of the $\phi $ scalar triplet are scalar DM candidates since their $Z_4$ charge corresponds to a nontrivial charge
under the preserved $Z_2$ symmetry. In addition, from the fermionic
assignments shown in Table \ref{tab:fermions}, it follows that $\Om _{nR}$
($n=1,2$) are fermionic DM candidates. The one-loop Feynman diagram
contributing to the $\mu $ parameter of the inverse seesaw is shown in
fig. \ref{Loopdiagramsmu}. Besides that, it is worth mentioning that the $%
Z_4$ discrete group allows preventing a tree-level charm quark mass, which
in the model only appears at one-loop level. On the other hand, the
appearance of tree-level up quark and SM down type quark masses is prevented
due to the $U(1)_X$ assignments of the particle spectrum. Such masses are
only generated at the one-loop level. The one-loop Feynman diagrams contributing
to the entries of the SM charged fermion mass matrices are shown in fig.\ref{Loopdiagramsf}. Notice that the electrically
charged scalars together with heavy vectorlike up (down) type quarks mediate one-loop level radiative
seesaw mechanisms that produce the masses for the down, strange and bottom quarks (up and charm quarks).
It is worth mentioning that the set of heavy vectorlike quarks $T_i$ ($i=1,2,3$), $B_n$ ($n=1,2$) is the
minimum amount of exotic quarks needed to provide one-loop level masses for the SM quarks lighter than
the top quark. In the case of minimal scalar content, one seesaw fermionic mediator is needed to provide one-loop
 level masses for each light SM fermions. Because of this reason, three vectors like-charged exotic leptons $E_i$ ($i=1,2,3$) are
 needed to provide one-loop level masses for the SM charged leptons. Having several fermionic seesaw mediators lower than
 the number of SM fermions lighter than the top quark will yield a proportionality between rows and (or) columns
of the SM fermion mass matrices yielding a vanishing determinant. \\
Furthermore, the $SU(3)_L\times U(1)_X$ gauge
symmetry is also spontaneously broken down to the $SU(2)_L\times U(1)_Y$
symmetry, at the scale to the $\La_{int}>\mathcal{O}(10)$ TeV, by
the vacuum expectation value (VEV) of the third component of the $SU(3)_L$
scalar triplet $\chi $.\\ \newpage
\hspace{20cm}\newline
\hspace{20cm}\newline
\hspace{20cm}\newline
\hspace{20cm}\newline
\hspace{20cm}\newline
\hspace{20cm}\newline
\hspace{20cm}\newline
\hspace{20cm}\newline
We consider $\La_{int}>\mathcal{O}(10)$ TeV
to comply with collider \cite{Salazar:2015gxa} and flavor constraints \cite%
{CarcamoHernandez:2005ka,Martinez:2008jj,Buras:2013dea,Buras:2014yna,Buras:2012dp}.
The breaking of the $SU(2)_L\times U(1)_Y$ symmetry is triggered by the VEV of the second component of the $SU(3)_L$
scalar triplet $\rho $, whose CP even neutral component corresponds to the $126$ GeV SM like Higgs boson.
Consequently, the spontaneous symmetry breaking chain proceeds as follows:
\bea
&&\mathcal{G}=SU(3)_C\times SU(3)_L\times U(1)_X\times
U(1)_{L_g}\times Z_4{\xrightarrow{\La_{int}}}  \crn
&&\hspace{7mm}SU(3)_C\times SU(2)_L\times U(1)_Y\times
Z_2^{(L_g)}\times Z_2{\xrightarrow{v_\rho}}  \crn
&&\hspace{7mm}SU(3)_C\times U(1)_Q\times Z_2^{(L_g)}\times Z_2.
\label{pt1}
\eea%
The $SU(3)_L$ scalar triplets of this model are represented as:
\bea
\chi &=&%
\begin{pmatrix}
\chi _1^0 \\
\chi _2^- \\
\fr{1}{\sqrt{2}}(v_\chi+\xi _\chi\pm i\zeta _\chi) \\
\end{pmatrix}%
,  \label{Scalartriplets} \\
\phi &=&%
\begin{pmatrix}
\phi _1^{+} \\
\fr{1}{\sqrt{2}}(\xi _\phi\pm i\zeta _\phi) \\
\phi _{3}^{+}%
\end{pmatrix}%
,\hspace*{0.3cm}\rho =%
\begin{pmatrix}
\rho _1^{+} \\
\fr{1}{\sqrt{2}}(v_\rho+\xi _\rho\pm i\zeta _\rho) \\
\rho _{3}^{+}%
\end{pmatrix}%
.  \notag
\eea%
whereas the $SU(3)_L$ fermionic triplets and antitriplets take the form:
\bea
Q_{1L} &=&\left( u_1,d_1,J_1,\right) _L^T,\hspace*{0.3cm}%
Q_{nL}=\left( d_n,-u_n,J_n\right) _L^T,  \crn
L_{iL} &=&\left( \nu _i,\,l_i,\,\nu _i^{c}\right) _L^T,\hspace*{%
0.3cm}n=2,3,\hspace*{0.3cm}i=1,2,3.  \label{Fermiontriplets}
\eea%
 where the first two components of the $SU(3)_L$ fermionic triplets and
antitriplets correspond to the left-handed SM fermionic
fields.

With the particle spectrum and symmetries specified in Tables \ref%
{tab:scalars} and \ref{tab:fermions}, we find that the relevant quark and
lepton Yukawa interactions crucial to generate the observed SM fermion mass
and mixing pattern, are described by the terms:
\bea
-\mathcal{L}_Y^{(q)} &=&y_1^{(J)}\overline{Q}_{1L}\chi
J_{1R}+\sum_{n=2}^{3}y_n^{(J)}\overline{Q}_{nL}\chi ^{\ast }J_{nR}+m_T%
\overline{T}_{2L}T_{2R}  \crn
&&+\sum_{k=1,3}\left( y_{k}^{(T)}\overline{T}_{kL}\si  T_{kR}+x_{k}^{(J)}%
\overline{T}_{kL}\va J_{1R}\right)  \crn
&&+y_{B}\overline{B}_{1L}\si  B_{1R}+m_{B}\overline{B}_{2L}B_{2R}+x_{J}%
\overline{B}_{1L}\va ^{\ast }J_{2R}  \crn
&&+\sum_{i=1}^{3}y_i^{(u)}\overline{Q}_{3L}\rho ^{\ast }u_{iR}+x_2^{(T)}%
\overline{Q}_{2L}\phi ^{\ast }T_{2R}  \crn
&&+\sum_{k=1,3}\left( x_{k}^{(T)}\overline{Q}_{3L}\rho ^{\ast
}T_{kR}+\sum_{i=1}^{3}x_{ki}^{(d)}\overline{T}_{kL}\zeta
_1^{+}d_{iR}\right)  \crn
&&+\sum_{i=1}^{3}x_{2i}^{(d)}\overline{T}_{2L}\zeta _2^{+}d_{iR}+x_{B}%
\overline{Q}_{1L}\phi B_{2R}  \crn
&&+\sum_{n=1}^2\sum_{i=1}^{3}z_{ni}^{(u)}\overline{B}_{nL}\zeta
_n^{-}u_{iR}+H.c,  \label{Lyq}
\eea%
\bea
-\mathcal{L}_Y^{(l)} &=&\sum_{i,j=1}^{3}\left[ y_{ij}^{(N)}\overline{L}%
_{iL}\chi N_{jR}+x_{ij}^{(L)}\varepsilon _{abc}\overline{L}_{iL}^{a}\left(
L_{jL}^C\right) ^{b}\left( \rho ^{\ast }\right) ^{c}\right]  \crn
&&+\sum_{n=1}^2\left(\sum_{m=1}^2y_{nm}^{(\Om )}\Om _{nR}\overline{\Om
_{mR}^C}\va +\sum_{i=1}^{3}x_{in}^{(N)}N_{iR}\overline{\Om _{nR}^C%
}\eta \right)  \crn
&&+\sum_{i,j=1}^{3}\left(x_{ij}^{(E)}\overline{L}_{iL}\phi
E_{jR}+z_{ij}^{(l)}\overline{E}_{iL}\varrho l_{jR}\right)  \crn
&&+\sum_{i=1}^{3}(m_E)_i\overline{E}_{iL}E_{iR}+H.c.  \label{Lyl}
\eea%
After the model symmetries are spontaneously broken, the above 
Yukawa interactions generate the one-loop level entries for the SM charged fermion
mass matrices, as indicated by the Feynman diagrams of figure \ref%
	{Loopdiagramsf}.  In the SM quark sector, only the third row of the SM up
type quark mass matrix is generated at tree level from the first term of the
 fourth line of Eq. (\ref{Lyq}). Such a third row is associated with the top quark
mass which arises at tree level in 
the model. Furthermore, note that the terms between the first
and third lines of Eq. (\ref{Lyq}) give rise to tree-level exotic quark
masses. In addition, from the lepton Yukawa interactions of (\ref{Lyl}), it
follows that the charged exotic leptons $E_i$ ($i=1,2,3$) (which are assumed to be physical fields) and the
right-handed Majorana neutrinos $\Om _{nR}$ ($n=1,2$) get tree-level masses at the scales $m_E$ and $v_{\varphi}$, respectively,
	whereas the SM charged leptons masses as well as the mass terms for the
	gauge singlet neutral leptons $N_{iR}$ ($i=1,2,3$) appear at one-loop level. Furthermore, from the quark
Yukawa interactions, it follows that the exotic quarks $J_i$ ($i=1,2,3$), $T_2$, $B_2$ and $B_1$, $T_k$ ($k=1,3$) do acquire
tree-level  masses at the scales $v_{\chi}$, $m_T$, $m_B$ and $v_{\si }$, respectively. Given that we are considering such scales to 
fulfill $\La_{int}>\mathcal{O}(10)$ TeV, the exotic quark masses are larger than 	their LHC mass limit of $1.3$ TeV \cite{Aaboud:2018pii}, 
for exotic Yukawa couplings of order unity.
In addition, from the one loop  Feynman diagrams of figure \ref{Loopdiagramsf}, we find that the SM quark mass matrices take the form:
\begin{eqnarray}
M_{U} &=&\left(
\begin{array}{ccc}
\varepsilon _{11}^{(u)} & \varepsilon _{12}^{(u)} & \varepsilon _{13}^{(u)}
\\
\varepsilon _{21}^{(u)} & \varepsilon _{22}^{(u)} & \varepsilon _{23}^{(u)}
\\
y_{1}^{\left( u\right) } & y_{2}^{\left( u\right) } & y_{3}^{\left( u\right)
}%
\end{array}%
\right) \frac{v}{\sqrt{2}},\hspace{0.2cm}\hspace{0.2cm}i=1,2,3,  \notag \\
\varepsilon _{1i}^{(u)} &=&\frac{1}{16\pi ^{2}}\frac{\lambda _{\phi
^{\dagger }\chi \varphi ^{\ast }\zeta _{2}^{+}}x_{B}z_{2i}^{\left( u\right)
}v_{\chi }v_{\varphi }}{m_{B_{2}}v}C_{0}\left( \frac{m_{\phi _{1}^{+}}}{%
m_{B_{2}}},\frac{m_{\zeta _{2}^{-}}}{m_{B_{2}}}\right)  \notag \\
&=&\frac{1}{16\pi ^{2}}\frac{x_{B}z_{2i}^{\left( u\right) }A_{1}}{m_{B_{2}}}%
C_{0}\left( \frac{m_{\phi _{1}^{+}}}{m_{B_{2}}},\frac{m_{\zeta _{2}^{-}}}{%
m_{B_{2}}}\right)  \notag \\
\varepsilon _{2i}^{(u)} &=&\frac{1}{16\pi ^{2}}\frac{\lambda _{\rho
^{\dagger }\chi \varphi ^{\ast }\zeta _{1}^{+}}y_{2}^{\left( J\right)
}z_{1i}^{\left( u\right) }v_{\varphi }}{m_{J_{2}B_{1}}}C_{0}\left( \frac{%
m_{\chi _{2}^{+}}}{m_{J_{2}B_{1}}},\frac{m_{\zeta _{1}^{-}}}{m_{J_{2}B_{1}}}%
\right) ,  \notag \\
&=&\frac{1}{16\pi ^{2}}\frac{y_{2}^{\left( J\right) }z_{1i}^{\left( u\right)
}A_{2}}{m_{J_{2}B_{1}}}C_{0}\left( \frac{m_{\chi _{2}^{+}}}{m_{J_{2}B_{1}}},%
\frac{m_{\zeta _{1}^{-}}}{m_{J_{2}B_{1}}}\right)  \label{MU}
\end{eqnarray}
%
\begin{eqnarray}
M_{D} &=&\left(
\begin{array}{ccc}
\varepsilon _{11}^{(d)} & \varepsilon _{12}^{(d)} & \varepsilon _{13}^{(d)}
\\
\varepsilon _{21}^{(d)} & \varepsilon _{22}^{(d)} & \varepsilon _{23}^{(d)}
\\
\varepsilon _{31}^{(d)} & \varepsilon _{32}^{(d)} & \varepsilon _{33}^{(d)}%
\end{array}%
\right) \frac{v}{\sqrt{2}},\hspace{0.2cm}\hspace{0.2cm}i=1,2,3,  \notag \\
\varepsilon _{1i}^{(d)} &=&\frac{1}{16\pi ^{2}}\sum_{k=1,3}\frac{\lambda
_{\rho \chi ^{\dagger }\varphi \zeta _{1}^{-}}y_{1}^{\left( J\right)
}x_{ki}^{\left( d\right) }v_{\varphi }}{m_{J_{1}T_{k}}}C_{0}\left( \frac{%
m_{\chi _{2}^{-}}}{m_{J_{1}T_{k}}},\frac{m_{\zeta _{1}^{+}}}{m_{J_{1}T_{k}}}%
\right) ,  \notag \\
&=&\sum_{k=1,3}\frac{y_{1}^{\left( J\right) }x_{ki}^{\left( d\right) }A_{2}}{%
16\pi ^{2}m_{J_{1}T_{k}}}C_{0}\left( \frac{m_{\chi _{2}^{-}}}{m_{J_{1}T_{k}}}%
,\frac{m_{\zeta _{1}^{+}}}{m_{J_{1}T_{k}}}\right)  \notag \\
\varepsilon _{2i}^{(d)} &=&\frac{1}{16\pi ^{2}}\frac{\lambda _{\phi \chi
^{\dagger }\varphi \zeta _{2}^{-}}x_{2}^{\left( T\right) }x_{2i}^{\left(
d\right) }v_{\chi }v_{\varphi }}{m_{T_{2}}v}C_{0}\left( \frac{m_{\phi
_{1}^{-}}}{m_{T_{2}}},\frac{m_{\zeta _{2}^{+}}}{m_{T_{2}}}\right) ,  \notag
\\
&=&\frac{x_{2}^{\left( T\right) }x_{2i}^{\left( d\right) }A_{1}}{16\pi
^{2}m_{T_{2}}}C_{0}\left( \frac{m_{\phi _{1}^{-}}}{m_{T_{2}}},\frac{m_{\zeta
_{2}^{+}}}{m_{T_{2}}}\right)  \notag \\
\varepsilon _{3i}^{(d)} &=&\frac{1}{16\pi ^{2}}\sum_{k=1,3}\frac{\lambda
_{\rho \chi ^{^{\dagger }}\varphi \zeta _{2}^{-}}x_{k}^{\left( T\right)
}x_{ki}^{\left( d\right) }v_{\chi }v_{\varphi }}{m_{T_{k}}v}C_{0}\left(
\frac{m_{\rho _{1}^{-}}}{m_{T_{k}}},\frac{m_{\zeta _{1}^{+}}}{m_{T_{k}}}%
\right)  \notag \\
&=&\sum_{k=1,3}\frac{x_{k}^{\left( T\right) }x_{ki}^{\left( d\right) }A_{3}}{%
16\pi ^{2}m_{T_{k}}}C_{0}\left( \frac{m_{\rho _{1}^{-}}}{m_{T_{k}}},\frac{%
m_{\zeta _{1}^{+}}}{m_{T_{k}}}\right)  \label{MD}
\end{eqnarray}
where the following loop function has been introduced \cite{Arbelaez:2016mhg}:
\begin{equation}
C_{0}\left( \widehat{m}_{1},\widehat{m}_{2}\right) =\frac{\widehat{m}_{1}^{2}%
\widehat{m}_{2}^{2}\ln \left( \frac{\widehat{m}_{1}^{2}}{\widehat{m}_{2}^{2}}%
\right) -\widehat{m}_{1}^{2}\ln \widehat{m}_{1}^{2}+\widehat{m}_{2}^{2}\ln
\widehat{m}_{2}^{2}}{\left( 1-\widehat{m}_{1}^{2}\right) \left( 1-\widehat{m}%
_{2}^{2}\right) \left( \widehat{m}_{1}^{2}-\widehat{m}_{2}^{2}\right) }.
\label{loopfunction}
\end{equation}
The experimental values of the SM quark masses and CKM parameters \cite{Xing:2007fb,Bora:2012tx,Tanabashi:2018oca}:
\begin{eqnarray}
\label{eq:Qsector-observables}
&&m_{u}(MeV) = 1.45_{-0.45}^{+0.56},\hspace{3mm}
m_{d}(MeV) = 2.9_{-0.4}^{+0.5},\nonumber\\
&&m_{s}(MeV) = 57.7_{-15.7}^{+16.8},\hspace{3mm}m_{c}(MeV) = 635\pm 86, \nonumber\\
&&m_{t}(GeV) =172.1\pm 0.6\pm 0.9, \hspace{3mm}\nonumber
m_{b}(GeV) = 2.82_{-0.04}^{+0.09}, \hspace{3mm}\nonumber\\
\nonumber
&&\sin \theta _{12} = 0.2254,\hspace{3mm}\nonumber
\sin \theta _{23} = 0.0414, \hspace{3mm}\nonumber
\sin \theta _{13} = 0.00355,\nonumber\\
&&J = 2.96_{-0.16}^{+0.20}\times 10^{-5},
\end{eqnarray}
can be well reproduced for the following benchmark point:
\begin{eqnarray}
z_{11}^{\left( u\right) } &\simeq &0.1,\hspace{0.2cm}\hspace{0.2cm}\hspace{%
0.2cm}\hspace{0.2cm}z_{12}^{\left( u\right) }\simeq 2.39,\hspace{0.2cm}%
\hspace{0.2cm}\hspace{0.2cm}\hspace{0.2cm}z_{13}^{\left( u\right) }\simeq
0.1,  \notag \\
z_{21}^{\left( u\right) } &\simeq &0.44,\hspace{0.2cm}\hspace{0.2cm}\hspace{%
0.2cm}\hspace{0.2cm}z_{22}^{\left( u\right) }\simeq 0.1,\hspace{0.2cm}%
\hspace{0.2cm}\hspace{0.2cm}\hspace{0.2cm}z_{23}^{\left( u\right) }\simeq
0.1,  \notag \\
x_{11}^{\left( d\right) } &\simeq &0.69+1.53i,\hspace{0.2cm}\hspace{0.2cm}%
\hspace{0.2cm}\hspace{0.2cm}x_{12}^{\left( d\right) }\simeq 0.025-0.86i,
\notag \\
x_{21}^{\left( d\right) } &\simeq &-1.12-3.46i,\hspace{0.2cm}\hspace{0.2cm}%
\hspace{0.2cm}\hspace{0.2cm}x_{22}^{\left( d\right) }\simeq 0.18+3.53i,
\notag \\
x_{23}^{\left( d\right) } &\simeq &2.25-0.59i,\hspace{0.2cm}\hspace{0.2cm}%
\hspace{0.2cm}y_{1}^{\left( u\right) }\simeq -0.65,\hspace{0.2cm}\hspace{%
0.2cm}\hspace{0.2cm}x_{13}^{\left( d\right) }\simeq 0.85\hspace{0.2cm}
\notag \\
x_{31}^{\left( d\right) } &\simeq &-1.74-3.49i,\hspace{0.2cm}\hspace{0.2cm}%
\hspace{0.2cm}\hspace{0.2cm}x_{32}^{\left( d\right) }\simeq 0.08+3.43i,%
\hspace{0.2cm}\hspace{0.2cm}  \notag \\
x_{1}^{\left( T\right) } &\simeq &0.95,\hspace{0.2cm}\hspace{0.2cm}\hspace{%
0.2cm}\hspace{0.2cm}x_{2}^{\left( T\right) }\simeq -1.3,\hspace{0.2cm}%
\hspace{0.2cm}\hspace{0.2cm}\hspace{0.2cm}x_{3}^{\left( T\right) }\simeq
1.25,  \notag \\
\hspace{0.2cm}y_{1}^{\left( J\right) } &\simeq &0.12,\hspace{0.2cm}\hspace{%
0.2cm}\hspace{0.2cm}\hspace{0.2cm}y_{2}^{\left( J\right) }\simeq 1.54,%
\hspace{0.2cm}\hspace{0.2cm}\hspace{0.2cm}\hspace{0.2cm}x_{B}\simeq 0.57\\
A_{2} &\simeq &1.9\mbox{TeV},\hspace{0.2cm}\hspace{0.2cm}\hspace{0.2cm}%
\hspace{0.2cm}A_{3}\simeq 4\mbox{TeV},\hspace{0.2cm}\hspace{0.2cm}\hspace{%
0.2cm}\hspace{0.2cm}m_{B_{2}}\simeq 7.7\mbox{TeV},  \notag \\
m_{J_{2}B_{1}} &\simeq &7.7\mbox{TeV},\hspace{0.2cm}\hspace{0.2cm}\hspace{%
0.2cm}\hspace{0.2cm}m_{T_{1}}\simeq 26\mbox{TeV},\hspace{0.2cm}\hspace{0.2cm}%
\hspace{0.2cm}\hspace{0.2cm}m_{T_{2}}\simeq 6\mbox{TeV},  \notag \\
m_{JT_{1}} &\simeq &928\mbox{GeV},\hspace{0.2cm}\hspace{0.2cm}\hspace{0.2cm}%
\hspace{0.2cm}m_{JT_{3}}\simeq 500\mbox{GeV},\hspace{0.2cm}\hspace{0.2cm}
\notag \\
m_{\rho _{1}^{\pm }} &\simeq &100\mbox{GeV},\hspace{0.2cm}\hspace{0.2cm}%
\hspace{0.2cm}\hspace{0.2cm}x_{33}^{\left( d\right) }\simeq 0.51,\hspace{%
0.2cm}\hspace{0.2cm}\hspace{0.2cm}\hspace{0.2cm}y_{3}^{\left( u\right)
}\simeq -0.44,  \notag \\
m_{\zeta _{1}^{\pm }} &\simeq &5\mbox{TeV},\hspace{0.2cm}\hspace{0.2cm}%
\hspace{0.2cm}\hspace{0.2cm}m_{\zeta _{2}^{\pm }}\simeq 2.7\mbox{TeV},%
\hspace{0.2cm}\hspace{0.2cm}\hspace{0.2cm}\hspace{0.2cm}m_{\chi _{2}^{\pm
}}\simeq 5\mbox{TeV},  \notag \\
m_{\phi _{1}^{\pm }} &\simeq &5.5\mbox{TeV},\hspace{0.2cm}\hspace{0.2cm}%
\hspace{0.2cm}\hspace{0.2cm}m_{T_{3}}\simeq 32\mbox{TeV},\hspace{0.2cm}%
\hspace{0.2cm}\hspace{0.2cm}\hspace{0.2cm}y_{2}^{\left( u\right) }\simeq
-0.61.\notag
\label{benchmarkquarksector}
\end{eqnarray}

Note that we use the experimental values of the quark masses at the $M_{Z}$ scale, from Ref. \cite{Bora:2012tx},
which are similar to those in \cite{Xing:2007fb}. The experimental values of the CKM parameters are taken from
Ref. \cite{Tanabashi:2018oca}. 
Furthermore, from the neutrino Yukawa interactions of Eq. (\ref{Lyl}) we get
the following neutrino mass terms:
\be
-\mathcal{L}_{mass}^{(\nu) }=\fr{1}{2}\left(
\begin{array}{ccc}
\overline{\nu _L^C} & \overline{\nu _R} & \overline{N_R}%
\end{array}%
\right) M_\nu\left(
\begin{array}{c}
\nu _L \\
\nu _R^C \\
N_R^C%
\end{array}%
\right) +H.c,  \label{Lnu}
\ee%
where the full neutrino mass matrix is given by:
\be
M_\nu=%
\begin{pmatrix}
0_{3\times 3} & \fr{v_\rho}{2\sqrt{2}}z^{\left( L\right) } & 0_{3\times
3} \\
\fr{v_\rho}{2\sqrt{2}}\left( z^{\left( L\right) }\right) ^T &
0_{3\times 3} & \fr{v_\chi}{\sqrt{2}}y^{(N)} \\
0_{3\times 3} & \fr{v_\chi}{\sqrt{2}}\left( y^{(N)}\right) ^T & \mu%
\end{pmatrix}, \label{massneutrino}
\ee
with
\[
z^{\left( L\right) }=\left( x^{(L)}\right) ^\dagger -\left( x^{(L)}\right)
^{\ast },
\]
and the entries of the submatrix, which are generated at one-loop level from the
Feynman diagram of figure \ref{Loopdiagramsmu}, take the form \cite{Ma:2006km}:
\bea
\mu _{ij}&=&\sum_{n=1}^2\fr{x_{in}^{(N)}x_{jn}^{(N)}\left(m_{\Om}\right)_{nm}}{16\pi
^2}\times\nn\\
&&\left[ \fr{m_{\eta _R}^2}{m_{\eta _R}^2-\left(m_{\Om}\right)_{nm}^2}%
\ln \left( \fr{m_{\eta _R}^2}{\left(m_{\Om}\right)_{nm}^2}\right)\right.\crn
&&\left. -\fr{m_{\eta _I}^2}{m_{\eta _I}^2-\left(m_{\Om}\right)_{nm}^2}\ln \left( \fr{m_{\eta _I}^2}{\left(m_{\Om}\right)_{nm}^2}\right) \right],\hs
\label{smallmu}\eea
where $i,j=1,2,3$, $\left(m_{\Om}\right)_{nm}=y_{nm}^{(\Om )}v_\va $ ($n,m=1,2$), $m_{\eta_R}=m_{Re\eta }$, $m_{\eta _I}=m_{Im\eta }$.

We would like to note that the neutrinos $\nu_{aL}, \nu_{aR}$ carry the lepton number one unit and the fields $N_{aR}$
have an opposite lepton number with respect to that of neutrinos. The Majorana mass term of $N_{aR}$, $\mu$-parameter, does not conserve the lepton number.
The smallness of $\mu$ can be naturally realized via a radiative correction given in (\ref{smallmu}). For another one-loop level realization of
 the inverse seesaw mechanism see for instance \cite{Dev:2012sg}.

Diagonalization of the neutrino mass matrix (\ref{massneutrino}) for $\mu,\fr{v_\rho}{2\sqrt{2}}z^{\left( L\right) }
 \ll \fr{v_\chi}{\sqrt{2}}y^{(N)}$ leads to the physical neutrino mass matrices as follows
\bea
M_\nu^{(1) } &\simeq &\left( \fr{v_\rho}{v_\chi}\right)
^2z^{\left( L\right) }\left( \left( y^{(N)}\right) ^T\right) ^{-1}\mu
\left( y^{(N)}\right) ^{-1}\left( z^{\left( L\right) }\right) ^T,
\label{Mnu1}\crn
M_\nu^{(2) } &\simeq &\fr{v_\chi}{2\sqrt{2}}\left(
y^{(N)}+\left( y^{(N)}\right) ^T\right) -\fr{1}{2}\mu ,\hspace{1cm}%
\hspace{1cm}  \label{Mnu2} \crn
M_\nu^{(3) } &\simeq &\fr{v_\chi}{2\sqrt{2}}\left(
y^{(N)}+\left( y^{(N)}\right) ^T\right) +\fr{1}{2}\mu ,  \label{Mnu3}
\eea%
where $M_\nu^{(1) }$ corresponds to the mass matrix for light active neutrinos ($\nu_{a}$),
 whereas $M_\nu^{(2) }$ and $M_{\nu}^{(3) }$ are the mass matrices for sterile neutrinos ($N_{a}^-, N_{a}^+$)
 which are superpositions of mostly $\nu_{aR}$ and $N_{aR}$ as $N_{a}^\pm \sim \fr{1}{\sqrt{2}} \left( \nu_{aR}\mp N_{aR}\right)$.
In the limit $\mu \rightarrow 0$, which corresponds to unbroken lepton number, we recover the massless
neutrinos of the SM. Due to the smallness of the $\mu$- parameter, the mass splitting of three pairs of sterile neutrinos becomes small,
 thus implying that the sterile neutrinos form 
pseudo-Dirac pairs and the light active neutrino mass scale resulting from the inverse seesaw mechanism takes the form:
\cite{Deppisch:2004fa}
\bea
\left( \fr{M_\nu^{(1)}}{0.1~\text{eV}}\right) =\left(\fr{\fr{v_\rho}{2\sqrt{2}}z_L^{(N)}}{100 ~\text{GeV}} \right)^2 \left(\fr{\mu}{1~ \text{keV}}\right)\left(\fr{M_{\nu}^{(2,3)}}{10^4 ~ \text{GeV}} \right)^{-2}.
\eea
Considering the neutrino Yukawa couplings, $x^{(N)}$,  of order unity, assuming for simplicity degenerate
Majorana neutrinos $\Om_{nR}$ ($n=1,2$) with a common mass of $20$ TeV, $m_{\eta_R}=400$ GeV,
with the splitting $m_{\eta _R}-m_{\eta _I}\simeq 0.63$ MeV, one sees that $\mu \simeq 1 \text{keV}$.
Thus, the light neutrino mass can be $0.1 \text{eV}$ if heavy neutrino mass
scale is a few ten $\text{TeV}$ or less and the Dirac neutrino mass is order of $100 \text{GeV}$. This means that this model can account for
 the smallness of the light active neutrino masses and several singlet pseudo-Dirac fermions have $O(1)$ TeV masses. As mentioned above, 
 the pseudo-Dirac fermions are degenerate in mass and thus they allow the implementation of the resonant leptogenesis mechanism crucial 
 to generate the baryon asymmetry of the universe (BAU) \cite{Pilaftsis:2005rv}. In the inverse seesaw mechanism, 
 the smallness of the $\mu$-parameter can naturally suppress the washout of pre-existing baryon
  asymmetry \cite{Blanchet:2009kk,Blanchet:2010kw}. However, to explain the smallness of the $\mu$-parameter, 
  the model introduces exotic fermions and scalar fields. They create new washout processes as
\bea lH^+ && \rightarrow N_{aR}\diagdown \nu_{aR} \rightarrow \eta \Om_{aR}, (E_{jR}, \phi_3^+),\crn
lH^+ && \rightarrow N_{aR} \rightarrow   \chi_1^0 \bar{\nu}_a,
\label{washout}\eea
where  $H^\pm \sim \rho_1^\pm \cos\al  +\chi_2^\pm \sin\al $. In the case, the baryogenesis temperature $(T_{BG})$ higher than
the inverse seesaw scale $\La_{ISS}$, the washout processes can only be avoided if the Yukawa couplings related to the processes
given in Eq.(\ref{washout}) must be too tiny. So the radiative inverse seesaw mechanism is insignificant.
The singlets $\Om_{aR}$ generate the small lepton number violation $\mu$-parameter via one-loop correction (see Eq.(\ref{smallmu})),
the baryogenesis occurs before electroweak symmetry breaking.
In the case $\La_{EM}< T_{BG}< \La_{ISS}$, if at least one particle has a mass lower than the value of the
baryogenesis temperature, the washout process can be suppressed \cite{Baldes:2013eva}. We assume that $y^{(N)}$ and $x^{(N)}$ are
 diagonal matrices and that there is an inverse hierarchical structure between them as $y^{(N)}_{33} \ll y^{(N)}_{22}, y^{(N)}_{11}$.
This choice allows us to successfully accommodate the experimental value of the electron and muon anomalous magnetic
 moments as indicated by Eq. (33). Therefore only the third generation of $N_a^\pm$ can give the contribution to the BAU.
We also assume that the masses of $E_{aR}, \Om_{aR}$ fields are larger than the masses of the lightest pseudo-Dirac
fermions $N_3^{\pm}\equiv N^{\pm}$, the lepton asymmetry parameter, which is induced by decay process of $N^{\pm}$, has the following form
\cite{Gu:2010xc,Pilaftsis:1997jf}
\bea
 \ep_{\pm}&&=\fr{\sum_{\al=1}^{3}\left\{\Ga(N_{\pm} \rightarrow l_\al H^+)-\Ga(N_{\pm} \rightarrow
 \bar{l}_\al H^-)\right\}}{\sum_{\al=1}^{3}\left\{\Ga(N_{\pm} \rightarrow l_\al H^+)+\Ga(N_{\pm}
 \rightarrow \bar{l}_\al H^-)\right \}} \nonumber \\ && \simeq \fr{Im \left(h^\dagger_\nu h_N h_\nu^\dagger h_N \right)_{33}}{8 \pi A_{\pm}}\fr{r}{r^2+\fr{\Ga^2_{\mp}}{m^2_{N^\mp}}},
\label{epss}\eea
where we have defined $h_\nu=-y^{(N)}\sin\al + z^{(L)}\cos\al , h_N=y^{(N)} \sin\al+z^{(L)}\cos\al $ , and
\bea
&& r \equiv \fr{m^2_{N^+}-m^2_{N^-}}{m_{N^+}m_{N^-}} \simeq \fr{2\mu}{\fr{v_\chi}{2\sqrt{2}}\left(y^N+(y^N)^T \right)},\\
&& A_{-}=(h_N^\dag h_N)_{33}, \hs  A_{+}=(h_\nu^\dag h_\nu)_{33}, \hs \Ga_{\pm}=\fr{A_{\pm}m_{N_1^\pm}}{8 \pi}. \nn
\eea
If one neglects the interference terms involving the two different sterile neutrinos 
$N^\pm$, the washout parameter $K_{N^+}+K_{N^-}$ is huge as mentioned in \cite{Dolan:2018qpy}.
However, the small mass splitting between the pseudo-Dirac neutrinos leads to a destructive interference in the scattering process \cite{Blanchet:2009kk}.
The washout parameter including the interference term is given as follows
\bea
K^{eff} \simeq \left ( K_{N^+} +K_{N^-
} \right )\de ^2,  \eea
with $\de  =\fr{m_{N^+}-m_{N^-}}{\Ga _{N^\pm}}, K_{N^\pm}= \fr{\Ga _\pm}{H}$ and $H(T)=\sqrt{\frac{4\pi^3g^{*}}{45}}\frac{T^2}{M_{P}}$, 
where $g^* =118$ is the number of effective relativistic degrees of freedom and $M_{Pl}=1.2\times 10^9$ GeV is the Planck constant.
In order to estimate the effective washout parameter, we use the extended Casas-Ibarra parametrization for the case
of the inverse seesaw mechanism \cite{Dolan:2018qpy}
\bea
z^{(L)} =\fr{2v_\chi}{v_\rho} \left(U_{PMNS} M_{\nu}^{\fr{1}{2}} R \mu^{\fr{-1}{2}} y^{(N)} \right),
\eea
where, $U_{PMNS}$ is the Pontecorvo-Maki-Nakagawa-Sakata (PMNS) leptonic mixing mass matrix, whose Standard parametrization is given by:
\be
\left(
\begin{array}{ccc}
c_{12}c_{13} & s_{12}c_{13} & s_{13}e^{-i \de } \\
-s_{12}c_{23}-c_{12}s_{23}s_{13}e^{i \de } & c_{12}c_{23}-s_{12}s_{23}s_{13}e^{-i \de } & s_{23}c_{13} \\
s_{12}s_{23}-c_{12}c_{23}s_{13}e^{i \de } & -c_{12}s_{23}-s_{12}c_{23}s_{13}e^{i \de } & c_{23} c_{13} \\
\end{array}
\right)
\label{UPMNS}
\ee
with $c_{ij}=\cos \theta_{ij}, s_{ij}=\sin \theta_{ij}$. Notice that, for the sake of simplicity, we dropped the Majorana phases in Eq. (\ref{UPMNS}).

\begin{figure}[tbp]
	\resizebox{8.5cm}{5cm}{\vspace{-2cm}%
		\includegraphics{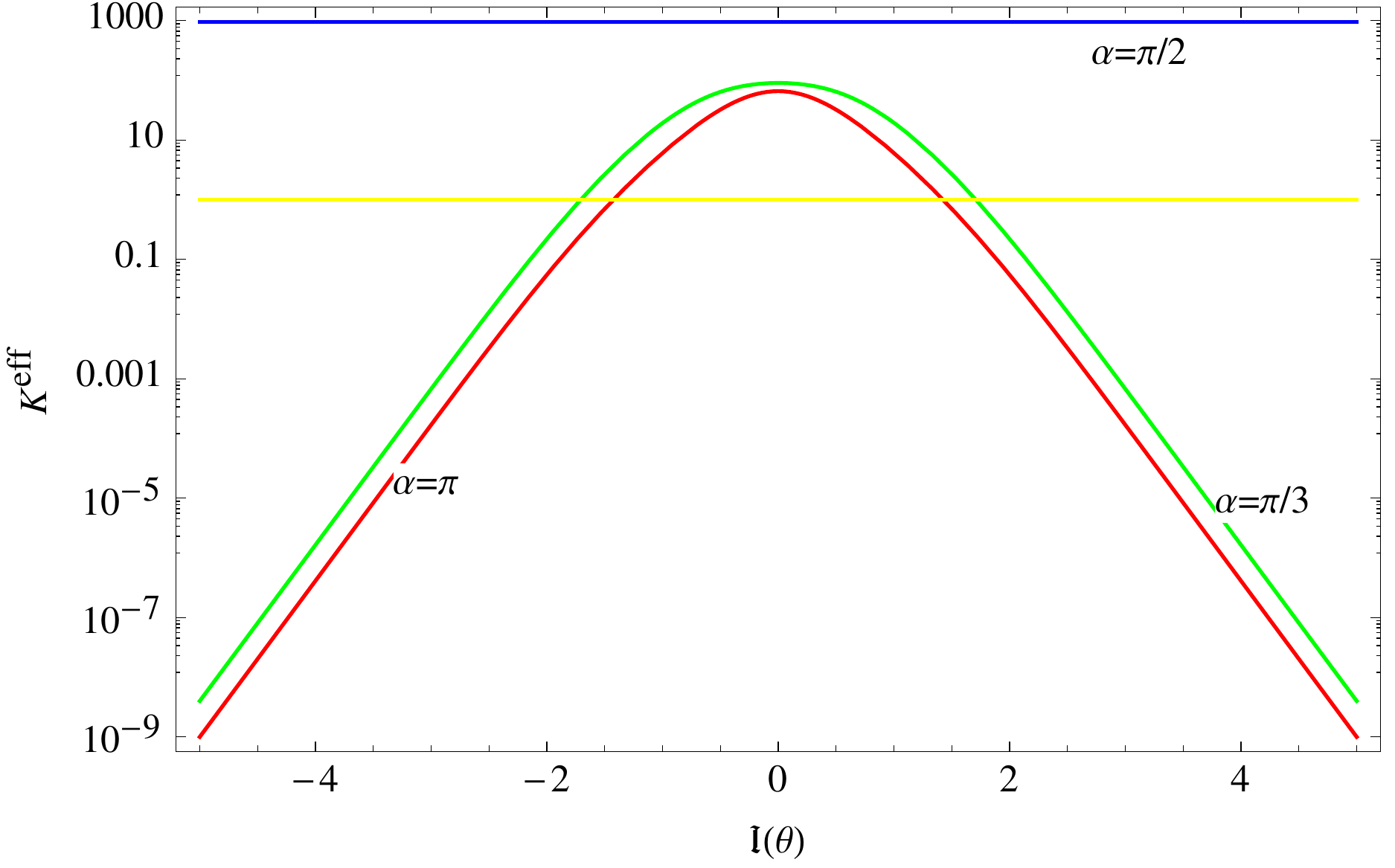}}\vspace{0cm}
	\caption{Effective washout parameter as a function of $\mathcal{I}[\theta]$ for different values of $\al $.}
	\label{Keff}
\end{figure}

For normal ordering, the current best fit values for
$\theta_{ij}$ and the CP violating phase are determined by \cite{deSalas:2017kay}
\bea
\frac{s_{12}^2}{10^{-1}}&& =3.20^{+0.20}_{-0.16}, \hs \frac{s_{23}^2}{10^{-1}}=5.47^{+0.20}_{-0.30}, \nonumber  \crn \frac{s_{13}^2}{10^{-2}} && = 2.160^{+0.083}_{-0.069}; \hs \frac{\de}{\pi}  =1.21^{+0.21}_{-0.15}. \nonumber
\eea
Furthermore, $M_\nu = diag(m_{\nu_1}, m_{\nu_2}, m_{\nu_3})$, being $m_{\nu_i}$ ($i=1,2,3$) the light active neutrino masses. The Dirac neutrino mass matrix is antisymmetric, thus implying that one light active neutrino is massless. Then, the light active neutrino masses for the case of normal hierarchy are given by: 
\bea
m_1=0, \hs m_2 = \sqrt{|\Delta m^2_{21}|},  \hs  m_3=\sqrt{|\Delta m^2_{31}|},
\eea
where the neutrino mass squared differences have the following experimental values \cite{deSalas:2017kay}:
\bea
&& \Delta m_{21}^2[10^{-5} eV^2]= 7.55^{+0.20}_{-0.16}, \nonumber \crn && \Delta m_{31}^2[10^{-3} eV^2]= 2.50\pm 0.03.\nonumber
\eea
Besides that $R$ is a complex orthogonal matrix which can be parametrized by
	\be
	R=\left(
	\begin{array}{ccc}
	c_yc_z & -s_xc_zs_y-c_xs_z & s_xs_z-c_xs_yc_z \\
	c_y s_z & c_x c_z- s_x s_y s_z & -c_z s_x-c_xs_ys_z \\
	s_y & s_xc_y & c_x c_y
	\end{array}
	\right),
	\ee
where $c_{x} = \cos x, s_x =\sin x$ and so on, being $x,y,z \in C $. For simplicity, we assume $x=y=z = \theta =\mathcal{R}[\theta]+ i \mathcal{I}[\theta]$
and we work on the basis where $y^{(N)}= Diag (y^{(N)}_1, y^{(N)}_2, y^{(N)}_3)$ is a diagonal matrix.
In the Fig.(\ref{Keff}), we plot the washout parameter as a function of $\mathcal{I}[\theta]$ for 
$\mu=1 \text{keV}, v_\chi=10^5\text{GeV},
v_\rho=246 \text{GeV}$ , and $\mathcal{R}[\theta]=1.2\pi$, $y_{11}^{\left( N\right) } \simeq 0.9$, $y_{22}^{\left( N\right) }\simeq 0.5i$ and  $y^{(N)}_3=0.01$.

The washout parameter depends on the imaginary part of complex angle $\theta$ and the mixing angle $\al$ (see in Fig.(\ref{Keff})). For $\al = \fr{\pi}{2}$,
 we predict an extremely high value of the washout parameter: $K_{eff}  \simeq 10^3$.

For $\al \neq \fr{\pi}{2}$, the model predicts both weak and strong washout regions. Especially, the $\theta$ angle is real, the washout parameter 
is larger than unit for any value of $\al $.
In the weak and strong washout region, the baryon asymmetry is related to the lepton asymmetry \cite{Pilaftsis:1997jf} as follows
\bea
\eta_b && =\fr{n_B}{s}=-\fr{28}{79}\fr{\ep_{\pm}}{g^*}, \hs \text {for} \hs K^{eff} \ll 1 , \\
\eta_b && =\fr{n_B}{s}=-\fr{28}{79} \fr{0.3\ep_{\pm} }{g^* K^{eff} \left(\ln K^{eff} \right)^{0.6}}, \hs \text {for} \hs  K^{eff} \gg 1, \label{etab1}\nn
\eea

\begin{figure}[tbp]
	\resizebox{8.5cm}{5cm}{\vspace{-2cm}%
		\includegraphics{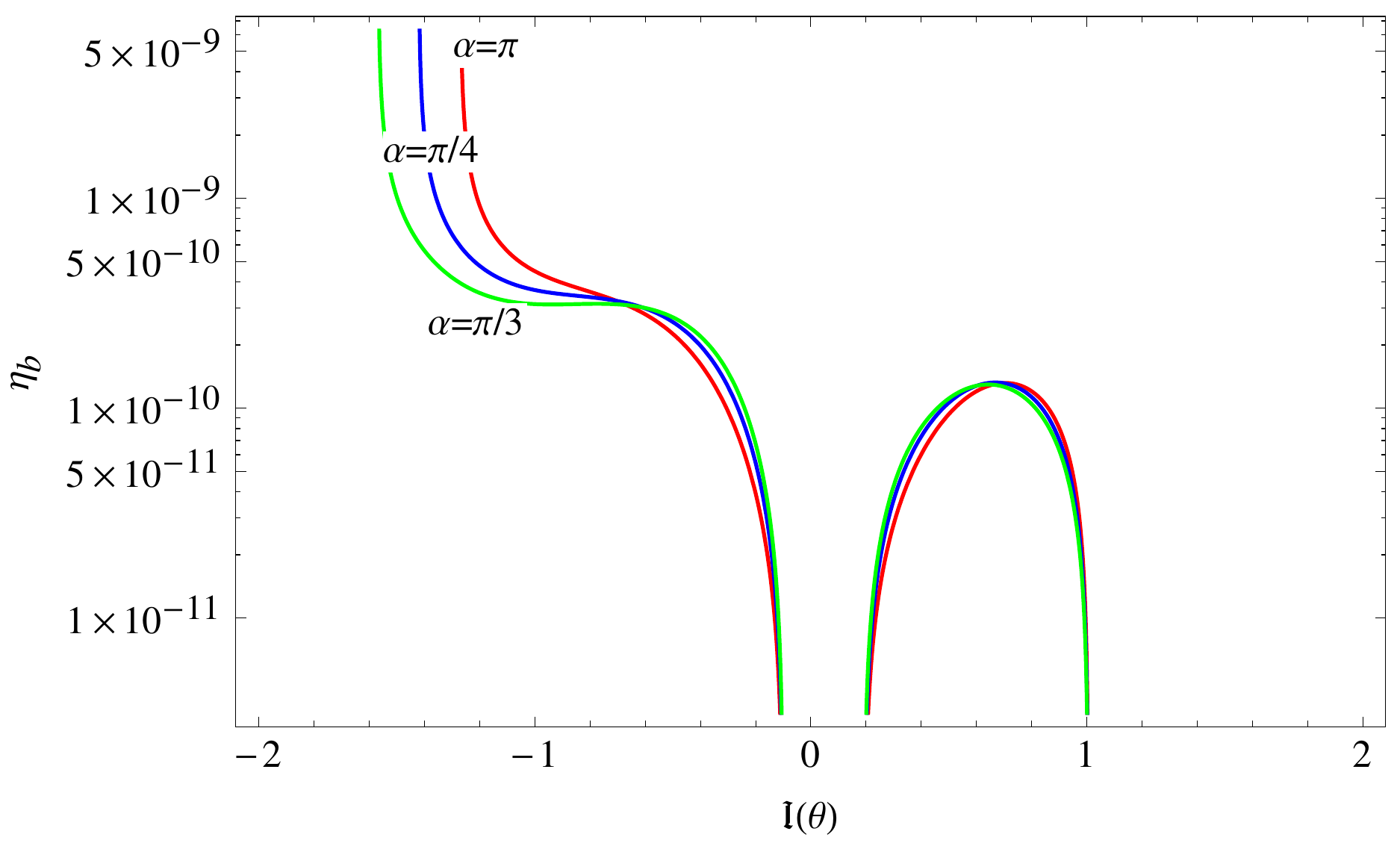}}\vspace{0cm}
\caption{Baryon asymmetry as a function of $\mathcal{I}[\theta]$ for fixing the mixing angle $\al$ in the strong washout region. }
	\label{Abaryon}
\end{figure}
\begin{figure}[tbp]
	\resizebox{8.5cm}{5cm}{\vspace{-2cm}%
		\includegraphics{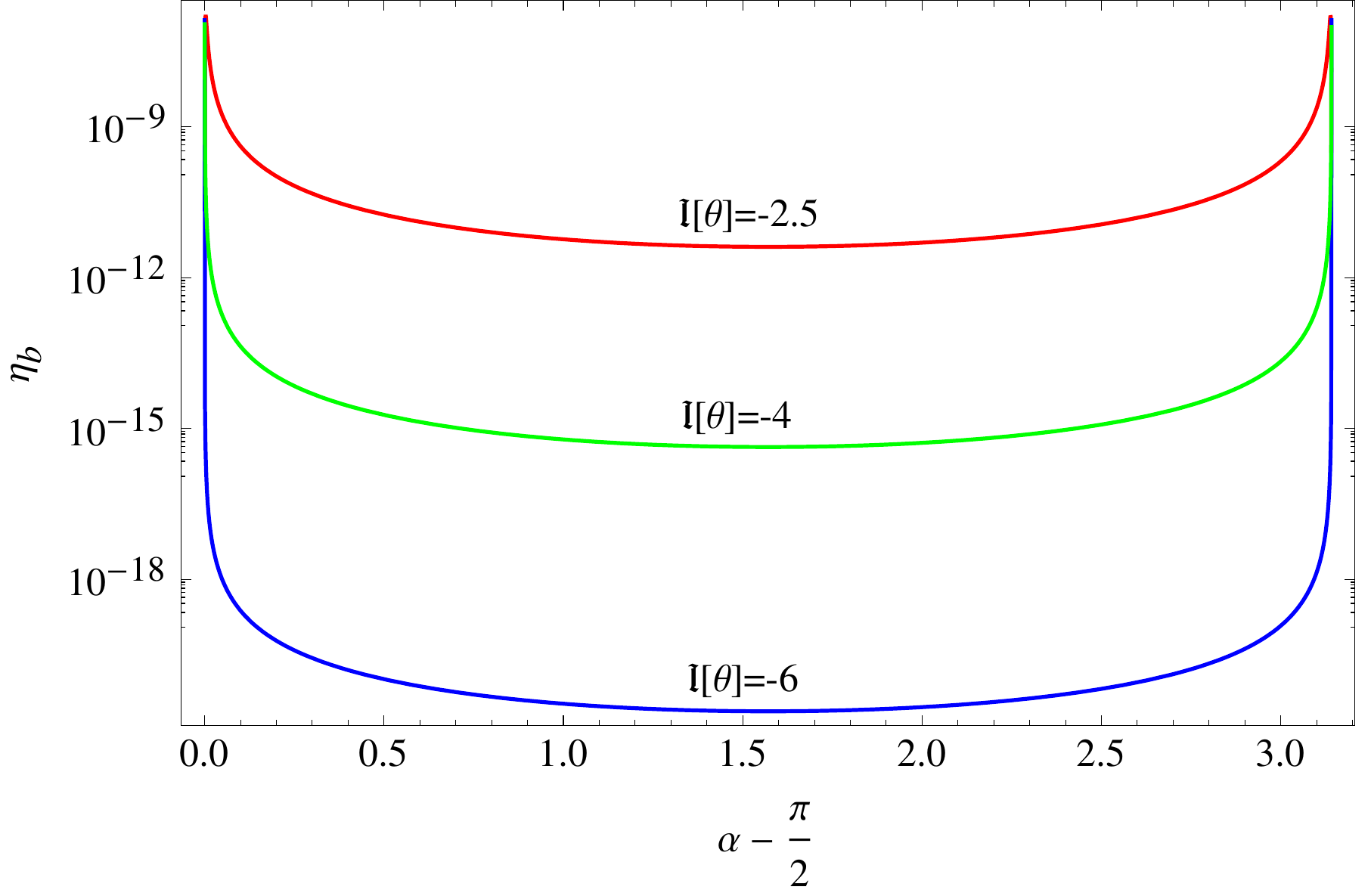}}\vspace{0cm}
	\caption{Baryon asymmetry as a function of the mixing angle $\al$ in the weak washout region}
	\label{weak}
\end{figure}

Fig.(\ref{Abaryon}) shows the baryon asymmetry $\eta_b$ as a function of $\mathcal{I}[\theta]$ for the difference choice $\al$ in the strong washout region. 
The amount of baryon asymmetry can reach its experimental value 
$\eta_B=6.2 \times 10^{-10}$. The value of $\mathcal{I}[\theta]$ that allows solving the baryon asymmetry depends on the value of the
 mixing angle $\al$. 
From results shown in the Figs. (\ref{Abaryon}), we can realize:
 \begin{itemize}
 	\item The amount of baryon asymmetry strongly depends on the mixing angle $\al$.

 	\item If $\theta$ is a real number, the predicted results are consistent with the results given in \cite{Dolan:2018qpy}.
 	
 \end{itemize}	
As $\al$ approaches $\frac{\pi}{2}$, one gets into the strong washout regime, the amount of baryon asymmetry suppressed.
Fig. (\ref{weak}) displays the amount of baryon asymmetry in the weak washout region via using a logarithmic scale 
for numerical data over a range of $\al$ value. At $\al= \frac{\pi}{2}$ or $\al=\frac{3\pi}{2}$, the curves are pulled because the suppression 
of the lepton asymmetry parameter. From the results shown in Fig. (\ref{weak}), one can find values of $\mathcal{I}[\theta]$ and $\al $ that allows
 creating a sufficient amount of baryon asymmetry consistent with its experimental value.\newpage


On the other hand, given that we are considering the exotic charged leptons $E_1$, $E_2$ and $E_3$ as physical fields,
 the SM charged lepton mass matrix is diagonal and the lepton mixings only arise from the neutrino sector.
 Consequently, to reproduce the experimental values of the neutrino mass squared splittings, leptonic mixing angles,
 and the leptonic Dirac CP-violating phase, for the scenario of normal neutrino mass hierarchy,
 we obtain the following numerical solution for the light active neutrino mass matrix:
 	\begin{widetext}
\be
M_\nu^{(1) }=\left(
\begin{array}{ccc}
	0.00298903\, -0.00104607 i & -0.00131313+0.00347814 i & -0.00661584+0.00316521 i \\
	-0.00131313+0.00347814 i & 0.0299126\, +0.000395314 i & 0.0213737\, -5.363173174949189\times 10^{-6} i \\
	-0.00661584+0.00316521 i & 0.0213737\, -5.363173174949189\times 10^{-6} i & 0.0249309\, -0.000337142 i \\
\end{array}
\right)eV
\ee
\end{widetext}
It is worth mentioning that such light active neutrino mass matrix generated from the one-loop level inverse seesaw mechanism allows to generate enough baryon asymmetry. Furthermore, from the light active neutrino mass matrix given above, we obtain that the effective Majorana neutrino mass parameter of neutrinoless double beta decay takes the value of around $3.2$ meV, which is beyond the reach of the present and forthcoming $0\nu\beta\beta$-decay experiments.

It is worth mentioning that the chosen benchmark of diagonal SM charged lepton mass matrix as well as diagonal $y^{(N)}$ matrix allows to have very 
suppressed charged lepton flavor violating (CLFV) processes.
	Note that the CLFV process $\mu\to e\gamma$ receives one loop-level contributions arising from vertex diagrams involving the exchange 
of heavy $W^{\prime}$ and the heavy sterile neutrinos. Given that we are considering the case of diagonal $y^{(N)}$ matrix, the off-diagonal
 entries of the sterile neutrino mass matrices will be very small compared with the diagonal ones, as follows from Eq. (\ref{Mnu3}), thus
  implying that the $\mu\to e\gamma$ will be suppressed by the square of the off-diagonal entries of the sterile neutrino mass matrices. 
  Thus, our model fulfill the constraints arising from CLFV processes.

On the other hand, as previously mentioned, the model has several DM candidates. Here we
assume that the lightest of them is the gauge singlet scalar $\eta _I$. Such singlet scalar dark matter candidate $\eta _I$
 mainly annihilates into $WW$, $ZZ$, $t%
\overline{t}$, $b\overline{b}$ and $HH$ via the Higgs portal quartic scalar interaction $%
H^{2}\eta_I^{2}$ that arises from the $\la _{\eta\rho}\eta \eta
^{\dagger }\rho \rho ^{\dagger }$ term of the scalar potential. We have numerically checked that in order
to successfully reproduce the experimental value, $\Omega h^{2}=0.1198\pm 0.0026$ \cite%
{Ade:2015xua}, of the relic density, the mass $m_{\eta_{I}}$ of the scalar
field $\eta _{I}$\ has to be in the range $400$ GeV$\ \lesssim m_{\eta_{I}}\lesssim $ $800$ GeV, for a quartic scalar coupling $\la _{\eta\rho}$ in
the range $1\lesssim\la _{\eta\rho}\lesssim 1.5$. Notice that we have taken {$m_{\eta _I}\simeq m_{\eta _R}=400$} GeV,
which corresponds to the lower bound for the mass of the singlet scalar DM candidate, as shown in Ref \cite{Bernal:2017xat},
where the scalar DM candidate is also a gauge singlet as in our model. This bound is consistent with the current constraints
arising from LUX and PandaX-II experiments \cite{Akerib:2016vxi,Tan:2016zwf}. In what concerns DM-direct-detection prospects,
 the scalar DM candidate
would scatter off a nuclear target in a detector via Higgs boson exchange in
the $t $-channel, thus yielding a constraint on the coupling of the $%
\left( \rho ^{\dagger }\rho \right) \eta_{I}\eta_{I}$ interaction. The corresponding scattering cross section takes
the form \cite{Garcia-Cely:2015ysa}:
\be
\si _{DM-p,n} \simeq 5 \times 10^{-44} \la _{\eta\rho}^2 \left(\fr{1 TeV }{m_{DM} } \right)^2 cm^2
\ee
Using the experimental value of the dark matter relic density, we find that the scattering cross section is predicted around $7 \times 10^{-45} cm^2$,
which is consistent with direct detection limits from the LUX \cite{Akerib:2012ys} and XENON1T \cite{Aprile:2012zx} experiments.
 In addition, note that the Yukawa
terms $x_n^{(E)}\overline{L}_{nL}\phi E_{nR}$ and $z_n^{(l)}\overline{E}_{nL}\varrho l_{nR}$ ($%
n=1,2$), as well as the trilinear scalar interaction $A\rho \phi ^\dagger
\varrho ^\ast $, will give rise to one-loop level contributions to the
muon and electron anomalous magnetic moments. Such one-loop level
contributions to the $\left( g_{e,\mu }-2\right) $ will involve the exchange
of CP-even and CP-odd scalars as well as of the heavy charged exotic leptons
$E_n$ ($n=1,2$) running in the internal lines of the triangular loops. Besides that, the muon and electron anomalous magnetic moments also receive one
loop-level contributions arising from vertex diagrams involving the exchange
of the electrically charged scalar $\chi^{+}_2$, which couples to the
right-handed Majorana neutrinos $N_{sR}$ ($s=1$ for electron and $s=2$ for muon). Then, the electron and muon anomalous magnetic moments take the form:
\bea
&& \De a_{e,\mu}= \crn &&\fr{y^2_{1,2}m_{e,\mu }^2}{8\pi ^2}\left[ I^{(1,2)}_{S}\left(
m_{E_{1,2}},m_{H_1}\right) -I^{(1,2)}_{S}\left( m_{E_{1,2}},m_{H_2}\right) \right.\crn
&&+\left. I^{(1,2)}_{P}\left( m_{E_{1,2}},m_{A_1}\right) -I^{(1,2)}_{P}\left(
m_{E_{1,2}},m_{A_2}\right) \right] \sin \theta \cos \theta\nn\\
&&-\frac{\left( y_{11,22}^{\left( N\right) }\right) ^{2}m_{e,\mu }^{2}}{8\pi
^{2}m_{\chi _{2}^{\pm }}^{2}}J\left( \frac{m_{\mu _{11,22}}}{m_{e,\mu }},%
\frac{m_{\mu _{11,22}}}{m_{\chi _{2}^{\pm }}}\right),
\eea
where $H_1= \cos\theta_S\func{Re}\varrho+\sin\theta_S\xi_\phi$, $H_2=-\sin\theta_S\func{Re}\varrho+\cos\theta_S\xi_{\phi}$, $A_1=\cos\theta_P\func{Im}\varrho+\sin\theta_P\xi_{\phi}$, $A_2=-\sin\theta_P\func{Im}\varrho+\cos\theta_P\zeta_{\phi}$ and for the
sake of simplicity we have set $x_n^{(E)}=z_n^{(l)}=y_n$ ($n=1,2$), $\theta_S=\theta_P$. In addition, the loop integrals are
given by \cite{Ma:2001mr,Diaz:2002uk,Kelso:2014qka}:
\bea
&&I^{(1,2)}_{S\left( P\right) }\left( m_E,m\right)= \crn && \int_{0}^1\fr{x^2\left(
1-x\pm \fr{m_E}{m_{e,\mu}}\right)dx}{m_{e,\mu}^2x^2+\left(
m_E^2-m_{e,\mu}^2\right) x+m^2\left( 1-x\right) },\nn\\
&&J\left( \frac{m_{\mu _{11,22}}}{m_{e,\mu }},%
\frac{m_{\mu _{11,22}}}{m_{\chi _{2}^{\pm }}}\right)=\nn\\
&&\int_{0}^{1}dx\frac{P_{+}\left( x,\frac{m_{\mu _{11,22}}}{m_{e,\mu }}\right)
+P_{-}\left( x,\frac{m_{\mu _{11,22}}}{m_{e,\mu }}\right) }{\left( \frac{%
m_{\mu _{11,22}}}{m_{\chi _{2}^{\pm }}}\right) ^{2}\left( 1-x\right) \left[
1-\left( \frac{m_{e,\mu }}{m_{\mu _{11,22}}}\right) ^{2}x\right]+x}
\eea
with
\bea
P_{\pm}(x,\epsilon)=-x\left(1-x\right)\left(x\pm\epsilon\right)
\eea
Notice that whereas the charged exotic vectorlike lepton $E_1$ contributes to the electron anomalous magnetic moment, $E_2$ contributes to the muon anomalous magnetic moment. This is due to the fact that we are considering the exotic charged leptons $E_1$ and $E_2$ as well as $E_3$ as physical fields, 
thus implying a diagonal SM charged lepton mass matrix, where the SM charged lepton masses are given by:
\begin{equation}
m_{i}=\frac{y_{i}^{2}}{2}\left[ F\left( m_{H_{1}},m_{A_{1}},m_{E_{i}}\right)-F\left(
m_{H_{2}},m_{A_{2}},m_{E_{i}}\right) \right]\sin 2\theta
\end{equation}
being $i=1,2,3$ and the loop function given by \cite{Ma:2006km}:
\begin{eqnarray}
  F\left( m_{1},m_{2},m_{3}\right)&=&\frac{m_{3}}{16\pi ^{2}}\left[ \frac{%
m_{1}^{2}}{m_{1}^{2}-m_{3}^{2}}\ln \left( \frac{m_{1}^{2}}{m_{3}^{2}}\right)\right.\nonumber\\
&-&\left.\frac{m_{2}^{2}}{m_{2}^{2}-m_{3}^{2}}\ln \left( \frac{m_{2}^{2}}{m_{3}^{2}}%
\right) \right] ,
\end{eqnarray}
From the previous relations we find that the SM charged lepton masses \cite{Xing:2019vks} and the muon anomalous magnetic 
moment can be very well reproduced for the following benchmark point:
\begin{eqnarray}
y_{1} &\simeq &0.43,\hspace{0.2cm}\hspace{0.2cm}\hspace{0.2cm}\hspace{0.2cm}%
y_{2}\simeq 1.01,\hspace{0.2cm}\hspace{0.2cm}\hspace{0.2cm}\hspace{0.2cm}%
y_{3}\simeq 3.07,\hspace{0.8cm}\theta =\frac{\pi }{4},\notag \\
m_{E_{1}} &\simeq &5.6\times 10^{5}\mbox{TeV},\hspace{0.2cm}\hspace{0.2cm}%
\hspace{0.2cm}\hspace{0.2cm}m_{E_{2}}\simeq 1.8\times 10^{2}\mbox{TeV}, \\
m_{E_{3}} &\simeq &2.7\times 10^{3}\mbox{TeV},\hspace{0.2cm}\hspace{0.2cm}m_{H_{1}}\simeq 10\mbox{TeV},\hspace{0.2cm}\hspace{0.2cm}m_{H_{2}}\simeq 5.2\mbox{TeV},  \notag \\
m_{A_{1}} &\simeq &18\mbox{TeV},\hspace{0.2cm}\hspace{0.2cm}\hspace{0.2cm}%
\hspace{0.2cm}m_{A_{2}}\simeq 14.5\mbox{TeV},\hspace{0.2cm}\hspace{0.2cm}%
\hspace{0.2cm}\hspace{0.2cm}m_{\chi_{2}^{\pm }}\simeq 0.1\mbox{TeV},\notag \\
y_{11}^{\left( N\right) } &\simeq &1.1,\hspace{0.2cm}\hspace{0.2cm}\hspace{%
0.2cm}y_{22}^{\left( N\right) }\simeq 0.7i,\hspace{0.8cm}\left\vert \mu _{11}\right\vert  =\left\vert \mu _{22}\right\vert \simeq 0.1%
\mbox{keV}.\notag
\end{eqnarray}
Then, for such benchmark point we get the following values for the muon and electron anomalous magnetic moments:
\begin{equation}
\Delta a_{\mu }\simeq 22.5\times 10^{-10},\hspace{0.2cm}\hspace{0.2cm}\hspace{%
0.2cm}\hspace{0.2cm}\Delta a_{e}\simeq -1.6\times 10^{-13}
\end{equation}
This shows that the obtained values for the muon and electron anomalous magnetic moments are consistent within the $1\sigma$ and $2\sigma$ experimentally allowed ranges, respectively, with their corresponding experimental values \cite{Hagiwara:2011af,Nomura:2018lsx,Nomura:2018vfz,Parker:2018vye}:
\bea
\left(\De a_\mu\right) _{\exp }&=&\left( 26.1\pm 8\right) \times
10^{-10},\crn
(\De a_e)_{\text{exp}}&=&(-0.88\pm0.36)\times 10^{-12}.
\eea
\newpage
In conclusion, we have constructed a minimal renormalizable theory that successfully explains the number of SM fermion generations,
the electric charge quantization, the SM fermion mass hierarchy, the tiny values of the light active neutrino masses, the lepton and baryon asymmetry
of the Universe, the observed DM relic density as well as the muon and electron anomalous magnetic moments. In our model, the top quark,
as well as the exotic fermions, obtain tree-level masses, whereas the SM charged fermions lighter than the top quark get one-loop level masses. 
Besides that, the tiny light active neutrino masses are produced from a one-loop level inverse seesaw mechanism.\vspace{5cm}
\begin{widetext}
\section*{Acknowledgments}
This research has received funding from Fondecyt (Chile), Grants No.~1170803; ICP. No. 2020.02; the Vietnam National Foundation for Science and Technology Development (NAFOSTED) under grant number 103.01-2019.312. A.E.C.H is very grateful to the Institute of Physics, VAST for the warm hospitality and for financing his visit.
\end{widetext}
\newpage
\begin{widetext}
 
\end{widetext}
\end{document}